**Title:** Improving Subgroup Analysis Using Methods to Extend Inferences to Specific Target Populations

**Running head:** Improving subgroup estimates with weighting

**Authors:** Michael Webster-Clark,[1,2] Anthony A. Matthews,[3] Alan R. Ellis,[4] Alan C. Kinlaw,[2] and Robert W. Platt.[1]

**Affiliations:**

1. McGill University, Montreal, QC, Canada.

2. University of North Carolina at Chapel Hill, NC, USA

3. Unit of Epidemiology, Institute of Environmental Medicine, Karolinska Institutet, Stockholm, Sweden.

4. North Carolina State University, Raleigh, NC, USA

**Funding:** This work was supported by NIH R01AG056479-01.

**Data sharing:** Researchers can apply for access to data on select participants in the PRIME trial through the Project Datasphere™ portal. Analytic code available upon request.

**Word count:** 3991/4000

**Corresponding author:**

Michael Webster-Clark, Pharm D, PhD

Department of Medicine, Division of Clinical Epidemiology, McGill University

Montreal, QC H4A 3J1

Email: michael.webster-clark@mcgill.ca

Phone: 1 919 966 7433 Fax: 1 919 966 2089

**Abstract:**

Subgroup analyses are common in epidemiologic and clinical research. Unfortunately, restriction to subgroup members to test for heterogeneity can yield imprecise effect estimates. If the true effect differs between members and non-members due to different distributions of other measured effect measure modifiers (EMMs), leveraging data from non-members can improve the precision of subgroup effect estimates. We obtained data from the PRIME RCT of panitumumab in patients with metastatic colon and rectal cancer from Project Datasphere™ to demonstrate this method. We weighted non-Hispanic White patients to resemble Hispanic patients in measured potential EMMs (e.g., age, KRAS distribution, sex), combined Hispanic and weighted non-Hispanic White patients in one data set, and estimated 1-year differences in progression-free survival (PFS). We obtained percentile-based 95% confidence limits for this 1-year difference in PFS from 2,000 bootstraps. To show when the method is less helpful, we also reweighted male patients to resemble female patients and mutant-type KRAS (no treatment benefit) patients to resemble wild-type KRAS (treatment benefit) patients. The PRIME RCT included 795 non-Hispanic White and 42 Hispanic patients with complete data on EMMs. While the Hispanic-only analysis estimated a one-year PFS change of -17% (95% C.I. -45%, 8.8%) with panitumumab, the combined weighted estimate was more precise (-8.7%, 95% CI -22%, 5.3%) while differing from the full population estimate (1.0%, 95% CI: -5.9%, 7.5%). When targeting wild-type KRAS patients the combined weighted estimate incorrectly suggested no benefit (one-year PFS change: 0.9%, 95% CI: -6.0%, 7.2%). Methods to extend inferences from study populations to specific targets can improve the precision of estimates of subgroup effect estimates when their assumptions are met. Violations of those assumptions can lead to bias, however.

Subgroup analyses are an important aspect of epidemiologic research. Typically, investigators separate the study population based on some characteristic of interest (e.g., geographic region), estimate effects in those groups, then compare estimates.[1] If the estimates differ statistically or clinically, we may declare the characteristic defining subgroups an important effect measure modifier (EMM) (a.k.a. "effect moderation," "statistical interaction," or "heterogeneity of effect")[2-5] and rely on subgroup estimates. If the estimates are similar, the full-population treatment effect estimate is typically treated as the best guess at the subgroup effect. These subgroup analyses can cause false positives and false negatives due to small sample sizes and random chance[6,7] and those false positives and false negatives can result in medication over- and under-use. Even when we correctly identify effect measure modification, subgroup estimates can be so imprecise that they are not useful in understanding harms and benefits of interventions.

Extending inferences from study populations to well-specified target populations (i.e., "transporting" treatment effect estimates)[8-13] may generate more precise subgroup estimates. Such methods use covariate information to estimate treatment effects in specific target populations assuming treatment consistency between the original study population and the targets and that measured covariates suffice to ensure (1) no EMMs differ between study and target populations or (2) membership in study vs target populations is independent of the outcome. Earlier applications of these methods extended inferences from randomized trials like ACTG[14] or RE-LY[15] to external populations like people living with HIV in the US or US Medicare beneficiaries taking dabigatran, respectively. More recently, they were used to estimate treatment effects in specific sites of multi-site trials[16,17] distributed data networks,[18] and subgroups of a target population defined by clinically relevant covariates.[19]

Another logical use for these methods is bolstering the precision of estimates in a study subgroup of interest. Here, we describe how one would apply this approach to the Panitumumab Randomized Trial in Combination with Chemotherapy for Metastatic Colorectal Cancer to Determine Efficacy (PRIME, National Clinical Trial ID: NCT00364013),[20, 21] which randomized patients to receive panitumumab or placebo alongside standard adjuvant chemotherapy for metastatic colon or rectal cancer. We target one subgroup where the method may be appropriate (improving estimates in Hispanic participants using data from non-Hispanic White participants), one subgroup with limited potential benefit (improving estimates in female participants using data from male participants) and one where the assumptions needed to apply the method are definitely violated creating bias (improving estimates for participants with wild-type Kirsten rat sarcoma virus [KRAS] tumors using data from participants with mutant-type KRAS tumors).

**METHODS**

**Data**

PRIME was a phase III clinical trial (NCT identifier: NCT00364013) that randomized 1,183 participants with metastatic colon or rectal cancer and no previous chemotherapy treatment to receive the standard of care (12 cycles of oxaliplatin with 5-fluorouracil and leucovorin, a.k.a. FOLFOX) or FOLFOX with panitumumab added each cycle.[20, 21] Trial enrollment began in August 2006 and results were published in March 2014. Because of phase II studies suggesting beneficial effects in participants with wild-type KRAS analyses were stratified by tumor KRAS variant. In the final trial analysis,[20] the panitumumab-containing regimen was shown to have superior median progression free survival (PFS) and median all-cause mortality in wild-type

KRAS participants, while in mutant-type KRAS participants FOLFOX alone appeared superior. De-identified individual-level data on 935 PRIME trial participants are publicly available on Project Datasphere™.

**Effect measure modification**

*Effect measure modifiers (EMMs):* EMMs play an essential role in governing whether one population can be used to estimate the treatment effect in another.[3] Speaking generally, EMMs are variables across which the treatment effect measure of interest (e.g., risk difference, risk ratio, hazard ratio, etc.) varies. In mathematical terms, a given binary variable *M* acts as an EMM for the risk difference of a binary exposure *X* on a binary outcome *Y* if it satisfies the inequality:

$$\Pr(Y^{X=1}|M=1) - \Pr(Y^{X=0}|M=1) \neq \Pr(Y^{X=1}|M=0) - \Pr(Y^{X=0}|M=0) \ ,$$

where $\Pr(Y^{X=1}|M=1)$ refers to potential outcome if we assign participants with *M=1* to the *X=1* treatment. While we cannot directly observe this probability, randomization and the consistency assumption[22] that the outcomes observed in those with X=1 equal the potential outcomes if those patients had been assigned to X=1 allows us to (in expectation) exchange the probability of the outcome among those with *M=1* randomized to *X=1*, written as $\Pr(Y|X=1, M=1)$, with $\Pr(Y^{X=1}|M=1)$. Similar inequalities for the risk ratio or hazard ratio define modification of those measures.

*Illustrating effect measure modification:* Consider EMM in the context of the PRIME trial. In the trial, the progression free survival benefit of adding panitumumab to a regimen of oxaliplatin, 5-fluourouracil, and leucovorin varied depending on the genetic variant of a tumor gene known as the Kristen Rat Sarcoma (KRAS) gene. This gene can either be wild-type or mutant-type. Substituting these terms into equation 1 gives

$$\Pr(PFS^{Pmab+FOLFOX}|WildKRAS) - \Pr(PFS^{FOLFOX}|WildKRAS) \neq$$

$$\Pr(PFS^{Pmab+FOLFOX}|MutantKRAS) - \Pr(PFS^{FOLFOX}|MutantKRAS)$$

with "PFS" referring to 1-year progression free survival, "FOLFOX" referring to the combination regimen with oxaliplatin and "Pmab" referring to panitumumab. These two risk differences are not equal, meaning that KRAS variant type is an EMM for the risk-difference-scale effect of adding panitumumab to FOLFOX. Applying the risk difference estimate obtained from analyzing the overall population to these subgroups would result in biased estimates and potentially incorrect clinical decisions.

*Conditional EMMs (cEMMs):* It is possible to incorporate additional variables into the assessment of a potential EMM by *conditioning* on them before checking for effect measure modification.[23] A variable is a cEMM when effects vary across levels of the variable *even after accounting for other covariates.* In mathematical terms, a given binary variable $M$ acts as a conditional EMM for the risk difference of a binary exposure $X$ on a binary outcome $Y$ if it satisfies the following inequality under a set of covariates $\boldsymbol{L}$ (bolding $\boldsymbol{L}$ to represent the potential to include multiple covariates):

$$\Pr(Y^{X=1}|M=1, \boldsymbol{L}=\boldsymbol{l}) - \Pr(Y^{X=0}|M=1, \boldsymbol{L}=\boldsymbol{l}) \neq$$

$$\Pr(Y^{X=1}|M=0, \boldsymbol{L}=\boldsymbol{l}) - \Pr(Y^{X=0}|M=0, \boldsymbol{L}=\boldsymbol{l})$$

where $\boldsymbol{L}=\boldsymbol{l}$ refers to a given distribution of the covariates included in $\boldsymbol{L}$. To assess whether this inequality is true, we need to compare the probability of the outcome in those with $M=1$ and $M=0$ under a common distribution of $\boldsymbol{L}$ (achieved via weighting or modeling). Whether a particular variable is a cEMM for a particular effect measure varies depending on the specific variables in $\boldsymbol{L}$.

*Illustrating conditional effect measure modification:* Consider again the PRIME trial and the directed acyclic graph of causal relationships between panitumumab, one-year progression-free survival, age, sex, and tumor KRAS variant in **Figure 1**. Suppose that age also modifies the effect of receiving a panitumumab-containing regimen vs a FOLFOX-only treatment regimen, with those over 65 not experiencing any additional treatment benefit from panitumumab (regardless of KRAS variant). The association between age and progression-free survival persists even after conditioning on KRAS variant (because age has an arrow directly into progression free survival on the graph) so substituting these variables into the equation evaluating whether a variable is a cEMM gives:

$$\Pr(PFS^{Pmab+FOLFOX}|Age \geq 65, KRAS = kras) - \Pr(PFS^{FOLFOX}|Age \geq 65, KRAS = kras) \neq$$
$$\Pr(PFS^{Pmab+FOLFOX}|Age < 65, KRAS = kras) - \Pr(PFS^{FOLFOX}|Age < 65, KRAS = kras)$$

If these two quantities differ, it means that the treatment effect differs between age groups even after standardizing them to have the same distribution of wild vs mutant-type KRAS (e.g., the distribution of KRAS in the participants over the age of 65) represented by the inclusion of "$KRAS = kras$" in each term. Age will act both as an EMM overall and as a cEMM conditional on KRAS variant.

*When an EMM is not a cEMM:* Importantly, a variable can be an EMM overall **but not** a cEMM when conditioning on a specific set of variables. Based on **Figure 1**, for example, sex is associated with age (as the women in the study are older than the men). The modification from age "flows" through the association between sex and age, resulting in a different treatment effect in male and female participants.[24, 25] Because the only causal path from sex to progression free survival goes through age, however, conditioning on age "blocks" the influence of the differing age distributions and sex **will not** be a cEMM conditional on age. In mathematical terms, the following equations are both true:

$$\Pr(PFS^{Pmab+FOLFOX}|Female) - \Pr(PFS^{FOLFOX}|Female) \neq$$
$$\Pr(PFS^{Pmab+FOLFOX}|Male) - \Pr(PFS^{FOLFOX}|Male)$$

meaning that sex is an EMM and that the overall population estimator will yield a biased estimate of the treatment effect in subgroups defined by sex, **and**:

$$\Pr(PFS^{Pmab+FOLFOX}|Female, Age = age) - \Pr(PFS^{FOLFOX}|Female, Age = age) =$$
$$\Pr(PFS^{Pmab+FOLFOX}|Male, Age = age) - \Pr(PFS^{FOLFOX}|Male, Age = age)$$

meaning that sex **is not** a cEMM conditional on age and the treatment effect for male and female participants in **Figure 1** will be the same after accounting for the differing age distributions between the two groups. Accounting for age would allow us to improve the precision of our treatment effect estimate in female participants, using information from the male participants, and vice versa.

**Using methods for extending inferences to improve subgroup estimates**

Fortunately, there are methods to extend inferences from a source population to target populations while accounting for differences in variable distributions. These methods require three key assumptions: **consistency of exposure** (the exposure being applied to the populations must not vary), **conditional exchangeability** between populations (measured variables must include a minimally sufficient adjustment set), and **positivity** between populations (all covariate combinations relevant to the treatment effect must be present in the source population if present in the target population).[17] If these assumptions are met with the non-members as source population and the subgroup members as the target, we can apply these methods to data from subgroup non-members to estimate effects in subgroup members while accounting for a particular set of variables. If subgroup membership is not a cEMM conditional on the measured variables, this process will generate an unbiased estimate of the effect in subgroup members.

*Weighting:* Weighting is a straightforward way to balance the distribution of multiple variables at once between two populations.[26-28] Weighting can construct a "pseudopopulation" with the same distribution of covariates as subgroup members from subgroup non-members. **Figure 2** depicts weighting male participants to resemble female participants with respect to age. We 1) obtain a data set with members and non-members, 2) use a predictive model (e.g., multivariable logistic regression) to estimate the probability of being a subgroup member conditional on the variables necessary to prohibit membership from being a cEMM (e.g., age), and 3) assign subgroup non-members weights equal to their predicted odds of subgroup membership (i.e., odds weights or OW) and subgroup members weights of 1, resulting in two populations with similar covariate distributions. If subgroup membership is not a cEMM conditional on the covariates included when estimating the predicted probabilities, then analyzing the subgroup members and the weighted non-members estimates a single quantity: the treatment effect in subgroup members.

*Combining estimates:* When used in the context of subgroup analyses, weighting can be used to separately analyze the subgroup members and the weighted non-members, generating point estimates with separate variances. The estimate from the analysis using the members of the subgroup of interest (all of whom have weights of 1) equals the estimate from a typical stratified subgroup analysis. While this estimate and the weighted estimate from subgroup non-members can and should be reported separately, pooling the two together will yield more informative treatment effect estimates, just as in a traditional meta-analysis.[29]

The most straightforward approach to combining the estimates from the two separate analyses is treating the subgroup members and odds-weighted non-members as a single analytic population. Other more complex options include pooling the estimates from the two analyses using methods traditionally used in meta-analysis and using stabilized odds weights rather than odds weights to restore the subgroup non-member population to its original sample size.[30] These latter approaches have two downsides, however: first, they assume that the two groups have independent variances (which is not technically the case),[31] and second, it gives more weight to the more precise of the two estimates. If subgroup membership is not a cEMM conditional on measured variables, these approaches will be unbiased and more precise. If subgroup membership is a cEMM conditional on measured variables and there are fewer subgroup members than non-members, however, the point estimate will more biased than that obtained from treating the subgroup members and odds-weighted non-members as a single analytic population.

**Applied example**

*Rationale for selecting the PRIME trial:* We chose to use this trial as a demonstration for two reasons. First, as mentioned previously, detailed de-identified individual-level data for the PRIME trial were publicly available on Project Datasphere™. Second, the PRIME trial included at least one well-understood biologically plausible mechanism of treatment effect heterogeneity in the form of tumor KRAS variant. This known treatment effect heterogeneity provides the opportunity to explore settings where applying inference-extending methods to subgroup analyses may be beneficial (when subgroup membership is an EMM but not a cEMM), may be of limited benefit (when minimal or no effect measure modification is expected), or may generate biased estimates (when subgroup membership is a cEMM).

*Beneficial case-Targeting Hispanic participants:* In the first case, we used data from non-Hispanic White participants to improve the precision of effect estimates in Hispanic participants.

Hispanic participants and non-Hispanic White participants differed in terms of KRAS (a known EMM) and other potential EMMs. While there is known heterogeneity between Hispanic patients and non-Hispanic White patients in terms of cancer outcomes, much of this heterogeneity is hypothesized to result from poorer healthcare access and financial toxicity, both of which are ameliorated in the context of randomized trial participants.[32] The influence of the fact that Hispanic patients are often diagnosed at later cancer stages (which results in poorer outcomes) is reduced in the context of a randomized trial that only included patients with metastatic cancer and can further take into account performance status at the start of treatment. This represents one scenario where these methods may be helpful: differences between members and non-members of the subgroup in terms of important EMMs, but the impact of conditional effect measure modification by the subgroup-defining characteristic has been limited by the study eligibility criteria and measured covariates.

Project Datasphere™ includes treatment, covariate, and outcome data for 45 Hispanic participants and 795 non-Hispanic White participants from PRIME. While 40% (17/45) of the Hispanic participants had wild-type KRAS, 60% (479/795) of the non-Hispanic White participants had wild-type KRAS. Given the potent effect measure modification by KRAS genotype, the difference in the distribution of wild-type KRAS suggests that Hispanic participants would benefit less by the addition of panitumumab than non-Hispanic White participants. It is also worth accounting for other factors that could modify the treatment effect, including age over 65, colon vs rectal cancer, liver metastases, performance status (a measure of functioning in cancer patients), and sex.

We estimated the probability of being a Hispanic participant with multi-variable logistic regression conditional on all these variables. We then assigned non-Hispanic White participants weights equal to their covariate-conditional odds of being a Hispanic participant, with Hispanic participants receiving weights of 1. Finally, we estimated differences between the two treatment regimens in the probability of one-year progression-free survival A) in the combined population with no weights, B) in non-Hispanic White participants with no weights, C) in non-Hispanic White participants with odds weights, D) in Hispanic participants with no weights, and E) in the combined population using the weights. Limits for the 95% confidence intervals of each quantity were obtained from the 2.5$^{th}$ and 97.5$^{th}$ percentiles of 2000 bootstrap iterations. We also calculated confidence limit differences (CLDs) to make it straightforward to compare variance across different analyses.

*Limited benefit case-Targeting female participants:* In the second case, we used data from male participants to improve the precision of effect estimates in female participants. Unlike the first case, the female participant subgroup was large (limiting the potential for improved precision for the combined weighted estimate compared to the subgroup-only estimate) and the distribution of EMM was similar (limiting the potential for substantial change in estimates). The data included 538 male participants and 327 female participants. We repeated the analyses from Case #1, including KRAS variant, age greater than 65, colon vs rectal cancer, presence of liver metastases, and performance status as adjustment variables.

*Biased case-Targeting wild-type KRAS participants:* For the final case, we used data from participants with mutant-type KRAS to improve the precision of effect estimates in participants with wild-type KRAS. In this case, since the subgroup of interest is defined by KRAS variant (itself a cEMM no matter what) inference-extending methods will yield biased estimates. The data included 513 participants with wild-type KRAS and 352 participants with mutant-type KRAS. We once again repeated the analyses conducted in Case #1, including biologic sex, age greater than 65, colon vs rectal cancer, presence of liver metastases, and performance status as adjustment variables.

**RESULTS**

*Key demographic characteristics:* **Table 1** shows the distributions of demographic and clinical variables in Hispanic participants and non-Hispanic White participants before and after applying odds weights for non-Hispanic White participants. Wild-type vs mutant-type KRAS variant, the proportion of participants over age 65, and the proportion of participants who were female all differed between Hispanic and non-Hispanic White participants. Weighting the non-Hispanic white participants to resemble the Hispanic participants eliminated these differences. **Supplemental Table 1** and **Supplemental Table 2** provide analogous information for the analyses targeting female participants and wild-type KRAS participants; differences between the subgroup members and non-members in these cases were smaller and also almost eliminated by applying odds weights. In all cases, the weighted total N of the subgroup non-members became approximately equal to that of the subgroup members as expected of odds weights.

*Targeting Hispanic participants:* **Figure 3** shows the point estimates and 95% confidence limits for the difference in 1-year progression-free survival between participants in the panitumumab + FOLFOX arm compared to participants in the FOLFOX alone arm for the analyses targeting the subgroup of Hispanic participants. The estimate in only Hispanic participants was very imprecise (1-year PFS difference: -17%, 95% CI: -45%, 9.1%, CID: 54%) compared to the estimate in non-Hispanic White participants (1-year PFS: 1.7%, 95% CI: -5.1%, 8.7%, CID: 14%). After odds weighting, the estimate in non-Hispanic White participants was closer to the null but less precise (1-year PFS: -0.5%, 95% CI: -10%, 9.4%, CID: 20%). Relative to the estimate in Hispanic participants only, the estimate from the combined weighted sample was closer to the null and nearly twice as precise (1-year PFS: -9.1%, 95% CI: -23%, 5.3%, CID: 28%), though the decreased precision of this weighted estimate vs the combined estimate (1-year PFS: 0.6%, 95% CI: -5.9%, 7.5%, CID: 13%) means it is difficult to truly say whether this represents a change in bias.

*Other subgroups:* **Table 2** shows results of the odds-weighted analyses targeting each subgroup. When targeting female participants or those with wild-type KRAS, combined weighted results were very similar to those obtained in the crude overall analysis (initial overall crude estimates: 1.2%, point estimate targeting women: 0.7%, point estimate targeting wild-type KRAS: 0.9%) and precision did not change substantially vs the full population estimate (initial overall confidence limit difference: 13%, confidence limit difference targeting women: 14%, confidence limit difference targeting wild-type KRAS: 13%). The increase in precision vs the subgroup-only estimates was still substantial, however: CLDs in the weighted pooled analysis were 14% vs 21% for female participants and 13% vs 18% for wild-type KRAS participants. When targeting wild-type KRAS participants, the weighted combined results centered on the null (1-year PFS : 0.9%, 95% CI: -6.0%, 7.2%) which does not align with the clinical benefit observed in the original PRIME RCT (hazard ratio of 0.80; 95% CI, 0.66 to 0.97).[20]

**DISCUSSION**

We almost doubled the precision of the estimated effect of adding panitumumab to treatment regimens for Hispanic participants with metastatic colorectal cancer in the PRIME trial by reweighting the non-Hispanic White participants to resemble the Hispanic participants in terms of key potential EMMs. When estimating treatment effects in specific subgroups of participants, methods for extending inferences to external targets allow individuals outside the subgroup to contribute to subgroup analyses without assuming homogeneous treatment effects. That said, these methods can generate biased estimates when their assumptions are violated (e.g., when there is an imbalance in a cEMM between the two groups or when group membership itself acts as a cEMM conditional on measured variables) as in the example targeting participants with wild-type KRAS tumors.

This represents a new application of existing methods in the transportability space including work on causal meta-analysis estimating site-specific treatment effects in multi-site randomized trials,[16, 17] combining results across data partners in distributed networks,[18] and synthesizing evidence to estimate conditional average treatment effects.[33, 34] Indeed, the methods and assumptions are one and the same. However, two characteristics of single-site study data make this setting better-suited for the analyses described here. First, the ability to combine individual-level data from subgroup members and non-members makes implementing weighting methods (or other analytic approaches to standardization) easier and more straightforward than in distributed analyses. Second, differential missingness[35] and measurement error[36] are less likely between subgroups in a single study than between sites in a multi-site study or distributed data network.

These methods possess major limitations. First, as discussed previously, they work **only** when accounting for differences in measured covariates eliminates effect measure modification and the consistency, exchangeability, and positivity assumptions are met. In subgroups where this is not the case (e.g., the wild-type KRAS participants, or if socioeconomic factors modify the treatment effect in Hispanic patients through paths besides access to care, financial toxicity, and cancer stage) restricting the analysis to subgroup members themselves and accepting the limited precision is the only unbiased approach. The methods also rely on the assumptions of other model-based methods to achieve validity, including but not limited a lack of measurement error, the absence of selection bias resulting from differential loss-to-follow-up or missing data, and correctly specified models with no omitted variables.[37, 38] Ignoring the prevalence of wild-type KRAS when reweighting non-Hispanic White participants to Hispanic participants, for example, resulted in a weighted non-Hispanic White estimate on the other side of the null (1-year PFS: 1.3%, 95% CI: -8.2%, 10.0%, CID: 18%). Omitted variable bias is particularly problematic because identifying variables to condition on can be difficult and, as discussed previously, studies may lack data on key socioeconomic factors that are necessary to render important subgroups not cEMMs. The variables necessary also differ depending on the effect measure of interest (e.g., risk ratio vs risk difference vs odds ratio),[23] and the methods to identify those variables typically have low power, particularly in trials powered to detect main effects.[39, 40]

Our illustrative analyses of the PRIME trial were limited. While we focused on weighting to extend inferences to specific targets, other strategies like outcome modeling and doubly robust methodologies are also valid.[12, 41] Many of these alternative methods and estimators are covered in-depth with respect to multi-site trials.[16, 17] Doubly robust methods may be particularly useful for reducing bias from incorrect model specification.[42] We also focused on combining the two populations after applying odds weights, rather than using stabilized odds weights, inverse variance weighting, random effects meta-analysis, or a Bayesian approach. Using stabilized rather than unstabilized odds weights resulted in a combined weighted estimate much closer to the estimate in weighted non-Hispanic White participants (1-year PFS: -1.5%, 95% CI: -10.9%, 8.3%, CID: 0.19). Bayesian methods could directly incorporate uncertainty surrounding the validity of methodologic assumptions within their data[43] and whether assumptions are more likely to hold for some subgroup non-members than others. We also focused on a randomized trial with exchangeability in expectation between treatment arms, rather than an observational study that must account for confounders and EMMs simultaneously, and on an internal subgroup, rather than a subgroup that is part of an external target population.[19] The additional assumptions required in those contexts should be investigated fully.

Finally, we must note that this approach does not diminish the importance of pre-specifying subgroups of clinical interest, particularly when performing analyses for regulators.[44] To avoid p-hacking, researchers should pre-specify subgroup target populations, subgroup comparator populations, and covariates for the final weighting or outcome model (or some algorithm for selecting covariates).

**Conclusion**

Estimating treatment effects in subgroups is a key part of epidemiologic research. Methods extending inferences to defined target populations can improve the precision of subgroup treatment effect estimates under certain assumptions. When those assumptions are violated, methods extending inferences can generate bias and false precision.

**Declaration of Conflicting Interests:** The authors declare that there are no conflicts of interest.

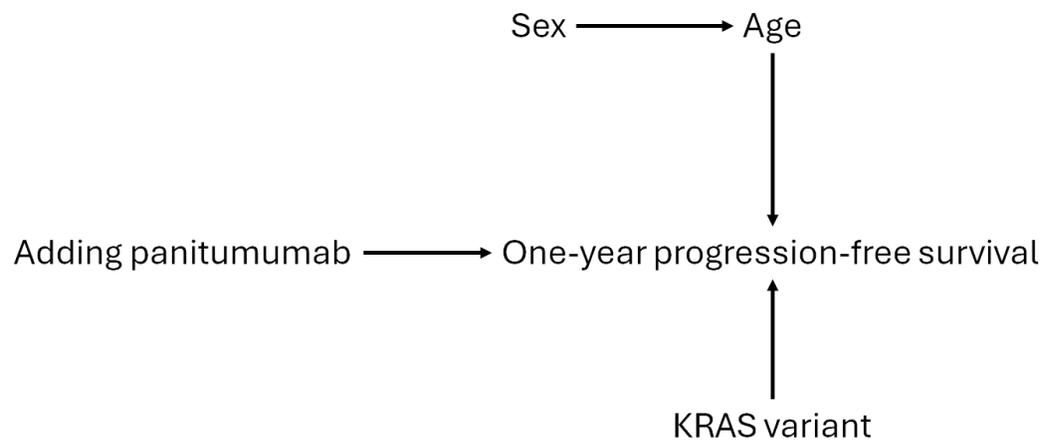

**Figure 1:** Example directed acyclic graph showing potential associations between the addition of panitumumab to chemotherapy regimens, one-year progression-free survival, type of KRAS variant, age, and sex in participants with metastatic colon or rectal cancer. Arrows represent causal effects of variables on one another.

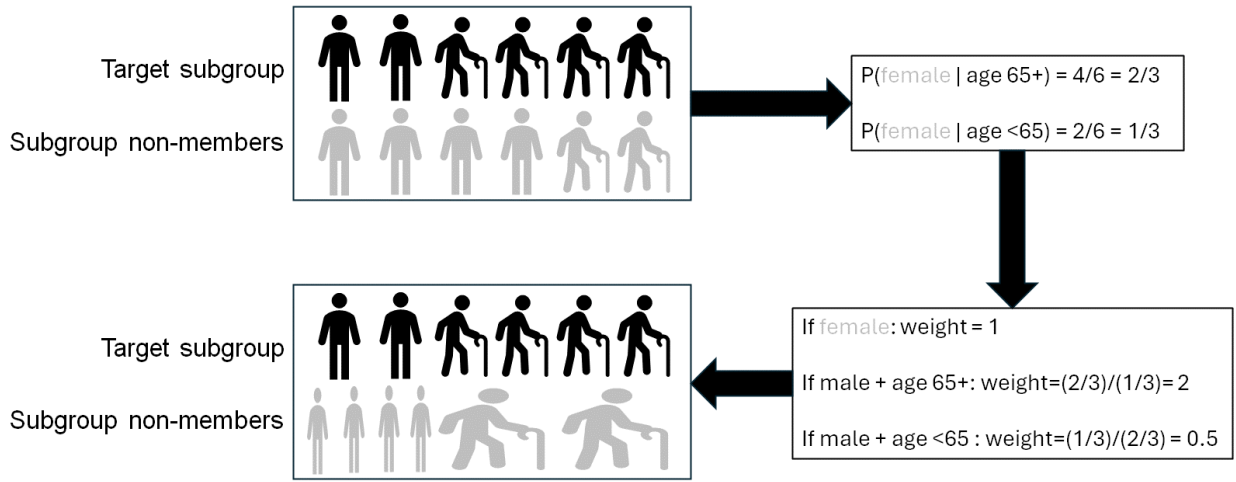

**Figure 2:** Graphic demonstrating weighting subgroup non-members to resemble subgroup members based on suspected effect measure modifiers, using an example reweighting male participants (in orange) to resemble female participants (in blue) in terms of frailty (represented with a cane).

**Table 1:** Distributions of key potential effect measure modifying covariates in Hispanic participants and non-Hispanic white participants before and after applying odds weights.

| Potential effect measure modifier | Non-Hispanic White participants (N=795) | Hispanic participants (N=42) | Odds-weighted non-Hispanic White participants (N=49.8) |
|---|---|---|---|
| Wild-type KRAS N (%) | 479 (60%) | 17 (40%) | 17.2 (41%) |
| Over age 65 N (%) | 312 (39%) | 12 (29%) | 12.0 (29%) |
| Female N (%) | 291 (37%) | 22 (52%) | 21.6 (52%) |
| Liver metastases N (%) | 704 (89%) | 39 (93%) | 38.6 (93%) |
| Colon cancer N (%) | 545 (69%) | 21 (50%) | 21.0 (50%) |
| ECOG 0 N (%) | 444 (56%) | 20 (47%) | 20.0 (48%) |
| ECOG 1 N (%) | 310 (39%) | 21 (50%) | 20.8 (50%) |
| ECOG 2 N (%) | 41 (5%) | 1 (2%) | 1.0 (2%) |

KRAS: Kristen rat sarcoma. ECOG: Eastern Cooperative Oncology Group.

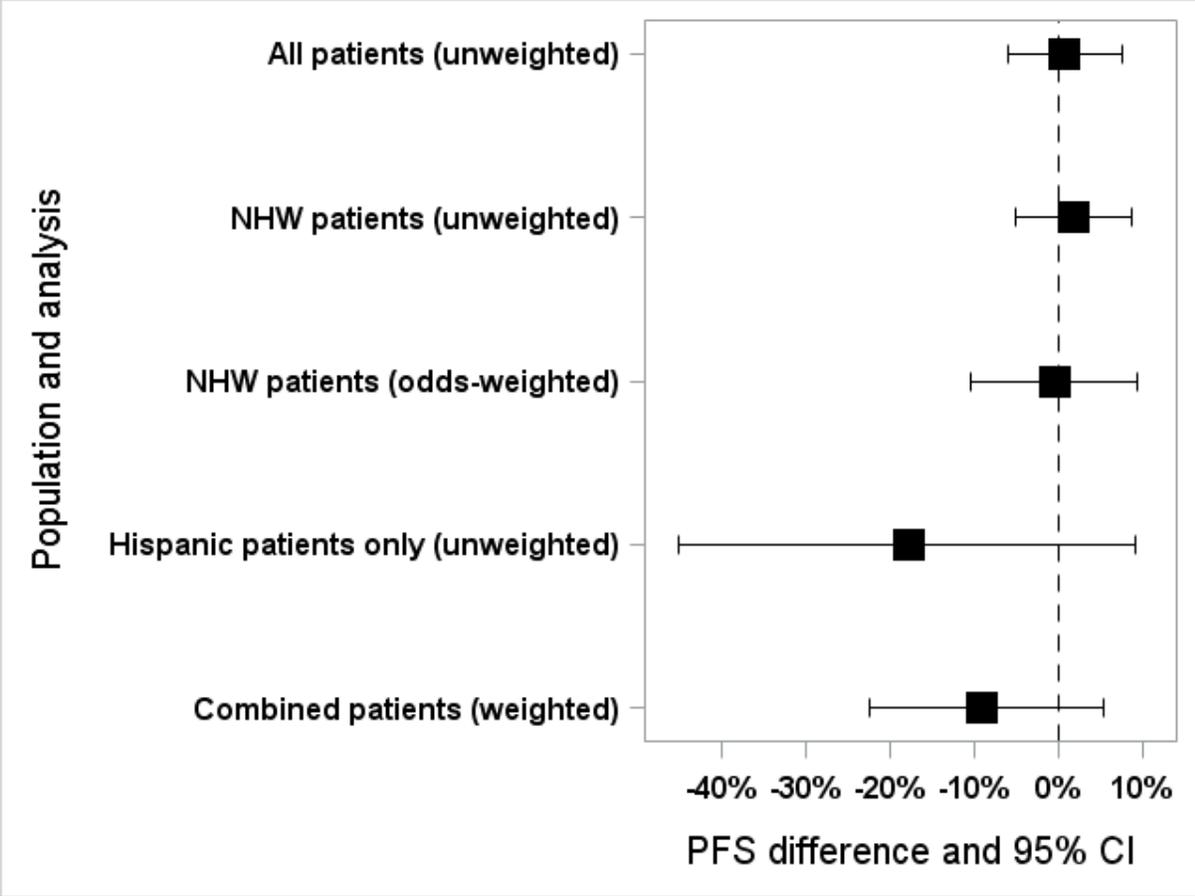

**Figure 3:** Estimates of the one-year progression free survival (PFS) difference analyzing only Hispanic participants, analyzing only non-Hispanic white (NHW) participants, and analyzing odds-weighted (OW) NHW participants, and results from a crude combined and weighted combined analysis. The 95% confidence intervals based on the 2.5th and 97.5th percentiles of 2,000 bootstrap iterations and the dashed line represents the null (PFS difference of 0).

**Table 2:** One-year progression free survival difference (PFSD) estimates from the unweighted combined analysis, the non-target subgroup members, the weighted non-target subgroup members, the target subgroup members, and the weighted combined analysis with 95% confidence intervals (CIs) from the percentiles of 2,000 bootstrap iterations.

| Analysis | Targeting Hispanic participant subgroup 1-year PFSD (95% CI) | CLD | Targeting female participant subgroup 1-year PFSD (95% CI) | CLD | Targeting wild-type KRAS subgroup 1-year PFSD (95% CI) | CLD |
| --- | --- | --- | --- | --- | --- | --- |
| Unweighted combined analysis[a] | 0.6% (-5.9%, 7.5%) | 0.13 | 1.2% (-5.3%, 7.6%) | 0.13 | 1.2% (-5.5%, 7.5%) | 0.13 |
| Non-target participants only | 1.7% (-10%, 9.4%) | 0.14 | 2.8% (-5.7%, 11%) | 0.17 | -6.5% (-16%, 2.8%) | 0.19 |
| Weighted non-target participants only | -0.5% (-10%, 9.4%) | 0.20 | 3.0% (5.9%, 11%) | 0.17 | -4.5% (-15%, 5.2%) | 0.20 |
| Target subgroup participants only | -17% (-45%, 9.1%) | 0.54 | -1.5% (-12%, 9.0%) | 0.21 | 6.2% (-2.9%, 15%) | 0.18 |
| Weighted combined analysis | -9.1% (-23%, 5.3%) | 0.28 | 0.7% (-6.2%, 7.5%) | 0.14 | 0.9% (-6.0%, 7.2%) | 0.13 |

PFSD: Progression free survival difference (positive = panitumumab beneficial)
[a]Unweighted combined analyses differ due to including participants that were neither Hispanic nor non-Hispanic White in the analyses when targeting female participants or participants with wild-type KRAS tumor variants.

Improving Subgroup Analysis Using Methods to Extend Inferences to Specific Target Populations

Supplemental Content

**Page 2-Supplemental Table 1:** Distributions of key potential effect measure modifying covariates in female participants and male participants before and after applying odds weights.

**Page 3-Supplemental Table 2:** Distributions of key potential effect measure modifying covariates in wild-type KRAS participants and mutant-type KRAS participants before and after applying odds weights.

**Supplemental Table 1:** Distributions of key potential effect measure modifying covariates in female participants and male participants before and after applying odds weights.

| Potential effect measure modifier | Male participants (N=538) | Female participants (N=327) | Odds-weighted male participants (N=326.6) |
|---|---|---|---|
| Wild-type KRAS N (%) | 329 (61%) | 184 (56%) | 183.1 (56%) |
| Over age 65 N (%) | 228 (42%) | 101 (31%) | 101.7 (31%) |
| Female N (%) | 0 (0%) | 327 (100%) | 0 (0%) |
| Liver metastases N (%) | 472 (88%) | 297 (91%) | 296.7 (91%) |
| Colon cancer N (%) | 348 (65%) | 233 (71%) | 232.9 (71%) |
| ECOG 0 N (%) | 289 (54%) | 192 (59%) | 189.3 (58%) |
| ECOG 1 N (%) | 219 (41%) | 122 (37%) | 123.2 (37.7%) |
| ECOG 2 N (%) | 30 (6%) | 13 (4%) | 14.0 (4%) |

**Supplemental Table 2:** Distributions of key potential effect measure modifying covariates in wild-type KRAS participants and mutant-type KRAS participants before and after applying odds weights.

| Potential effect measure modifier | Mutant-type KRAS participants (N=352) | Wild-type KRAS participants (N=513) | Odds-weighted mutant-type KRAS participants (N=511.9) |
|---|---|---|---|
| Wild-type KRAS N (%) | 352 (0%) | 513 (100%) | 0 (0%) |
| Over age 65 N (%) | 134 (38%) | 195 (38%) | 198.5 (39%) |
| Female N (%) | 143 (41%) | 184 (36%) | 186.3 (36%) |
| Liver metastases N (%) | 316 (90%) | 453 (88%) | 451.0 (88%) |
| Colon cancer N (%) | 255 (72%) | 326 (64%) | 327.6 (64%) |
| ECOG 0 N (%) | 184 (52%) | 297 (58%) | 296.8 (58%) |
| ECOG 1 N (%) | 153 (43%) | 188 (37%) | 189.7 (37%) |
| ECOG 2 N (%) | 15 (4%) | 28 (5%) | 25.4 (5.0%) |